\documentclass[aps,prd,preprintnumbers,showpacs,showkeys,nofootinbib,
superscriptaddress,fleqn,floatfix,nofootinbib,
notitlepage,tightenlines]{revtex4-2}  
\usepackage{amsmath,amsfonts,amssymb,amscd,amsxtra,amsthm}
\usepackage{graphicx}  % Include figure files
\usepackage{epstopdf}
\usepackage{dcolumn}  % Align table columns on decimal point 
\usepackage{bm}  
\usepackage{hyperref}      % bold math
\usepackage{slashed}
\usepackage{cancel} 
\usepackage{float}
\usepackage{mathtools} 
\usepackage{amsbsy}
\usepackage{amstext}

\usepackage[normalem]{ulem}% \sout{old text} for strikeout 
\usepackage[dvipsnames]{xcolor}% For blue in-text comments and
\usepackage[utf8]{inputenc} %raggedright - Each line of the caption
\usepackage{booktabs}
\usepackage[normalem]{ulem} % \sout{old text} for strikeout
\usepackage[dvipsnames]{xcolor} % For blue in-text comments and
\usepackage{tabularx}
\usepackage{enumitem}  
\usepackage{array} 
\usepackage{multirow}
\renewcommand\sout{\bgroup \color{red} \ULdepth=-.5ex \ULset}

\makeatletter

\begin{document}  
\preprint{INHA-NTG-08/2022}
\title{\Large Molecular nature of the $a_1$(1260) axial-vector meson}

\author{Samson Clymton}
\email[E-mail: ]{sclymton@inha.edu}
\affiliation{Department of Physics, Inha University,
Incheon 22212, Republic of Korea }

\author{Hyun-Chul Kim}
\email[E-mail: ]{hchkim@inha.ac.kr}
\affiliation{Department of Physics, Inha University,
Incheon 22212, Republic of Korea }
\affiliation{School of Physics, Korea Institute for Advanced Study 
  (KIAS), Seoul 02455, Republic of Korea}
%---------------------------------------------------
\begin{abstract}
We investigate $\pi\rho$ scattering based on the coupled-channel
formalism with the $\pi\rho$ and $K\bar{K}^*$
($\bar{K}K^*$) channels included. We construct the kernel amplitudes
by using the meson-exchange model and compute the coupled integral
equation for $\pi\rho$ scattering. By performing the partial-wave
expansion, we show explicitly that the $a_1(1260)$ meson is
dynamically generated by the coupled-channel formalism. The $a_1$
meson only appears by including the $K\bar{K}^*$ ($\bar{K}K^*$)
channel. We obtain the pole position of the $a_1$ meson as $\sqrt{s_R}
= (1170.7 - i 173.0)$ MeV. We conclude that the $a_1$ meson can be 
interpreted as a kaon and vector kaon molecular state. 
\end{abstract}
%---------------------------------------------------
\maketitle
\section{Introduction~\label{sec:1}} 
The $a_1(1260)$ meson is the first axial-vector
meson. If chiral symmetry $\mathrm{SU(2)\otimes SU(2)}$ is unbroken, 
then the $\rho$ and $a_1$ mesons form a chiral
doublet~\cite{Weinberg:1967kj, Gell-Mann:1960mvl} as the $\pi$ and 
$\sigma$ ($f_0(500)$) do. The existence of the $a_1$ meson has
been well established since the ACCMOR
Collaboration~\cite{ACCMOR:1979lrp} confirmed it in partial wave 
analyses of the $\pi^-\pi^-\pi^+$ system. However, the values of its
mass and width do not reach an experimental consensus. For
example, the ARGUS Collaboration at DESY~\cite{ARGUS:1986yqj} found  
the mass of the $a_1$ meson to be $m_{a_1}=(1\,046\pm 11)$ MeV and its
width to be $\Gamma_{a_1}=(521\pm 27)$ MeV in the decay $\tau\to
\pi^-\pi^-\pi^+ \nu_\tau$ whereas the CLEO Collaboration announced a
rather large values $m_{a_1}=(1\,331\pm 10\pm 3)$ MeV and
$\Gamma_{a_1}=(814\pm 36\pm 13$) MeV in the decay $\tau\to
\pi^-\pi^0\pi^0 \nu_\tau$~\cite{CLEO:1999rzk}. Recently, the LHCb
Collaboration measured $m_{a_1}=(1\,195.050 \pm 1.045 \pm 6.333$) MeV
and $\Gamma_{a_1}=(422.013 \pm 2.096 \pm 12.723)$ MeV in $D^0\to
K^{\mp} \pi^{\pm}\pi^\pm \pi^\mp$
decays~\cite{LHCb:2017swu}. Moreover, one should keep in mind that  
there is an inevitable model dependence in analyzing the experimental
data. The Particle Data Group (PDG) estimates the average values of
the $a_1$ mass and width as  $m_{a_1}=(1\,230\pm 40)$ MeV and
$\Gamma_{a_1}=(420\pm 35$) MeV. 
 
Since the $a_1$ meson has quantum numbers as
$J^{PC}=1^{++}$, it can be constructed as $q\bar{q}$
($1^3\mathrm{P}_1$) state. It is an isovector meson with negative $G$
parity. Dankowych et al.~\cite{Dankowych:1981ks}
carried out the isobar-model partial-wave analysis of high
statistics data on $\pi^-p\to \pi^+\pi^-\pi^0 n$ from the Argonne
National Laboratory zero-gradient synchrotron. They extracted the
partial-wave cross section of $\pi\rho$ scattering in
$\mathrm{S}$ and $\mathrm{D}$ waves.  The $a_1$ resonance was
observed in the $\mathrm{S}$-wave cross section with the broad width,
which reaches the kaon and vector kaon ($\bar{K}K^*$) threshold. It was
even seen that the $a_1$ meson decays into $\bar{K}$ and
$K^*$~\cite{CLEO:1999rzk, Belle:2002gzj, CLEO:2004hrb}. 
It implies that the $a_1$ meson may be strongly coupled to
the $K$ and $K^*$. Thus, the $a_1$ meson may contain the tetraquark 
component, or it can even be interpreted as the molecular
state~\cite{Basdevant:1977ya, Roca:2005nm, Lutz:2003fm,
  Nagahiro:2011jn}. A similar situation can be found in the case of the
scalar-isovector meson $a_0(980)$, which is often interpreted 
either as a tetraquark state or as a resonance appearing from the $\pi
\eta$ and $K\bar{K}$ coupled channels~\cite{Jaffe:1976ig,
  Achasov:1980tb, Baru:2003qq, Wang:2022vga}. The scalar mesons
$f_0(500)$ and $f_0(980)$ meson are also considered as the tetraquark
or molecular states~\cite{Lohse:1990ya, Lohse:1990ew, Baru:2003qq,
  Amsler:2004ps, Ahmed:2020kmp, Achasov:2020aun}. In particular, the
$f_0(980)$ is just below the $K\bar{K}$ threshold, it can be
regarded as a $K\bar{K}$ molecular state~\cite{Lohse:1990ew,
  Oller:1997ti}.  

Janssen et al. ~\cite{Janssen:1994uf} constructed the meson-exchange
model for $\pi\rho$ scattering with the effective Lagrangian,
including the $a_1$ meson explicitly. In the present work, we will
extend the work of Ref.~\cite{Janssen:1994uf} by considering the
coupled-channel formalism. We add the $K\bar{K}^*$ ($\bar{K} K^*)$
channel to the $\pi\rho$ channel but exclude the $a_1$ meson. 
We will show how the $K\bar{K}^*$ ($\bar{K} K^*)$ channel generates
dynamically the $a_1$ meson and describe successfully the S-wave cross
section. We first formulate the kernel amplitude based on the
meson-exchange model. We treat the vector meson based on the hidden
local gauge symmetry~\cite{Bando:1984ej, Bando:1985rf}. This has a
certain merit that the coupling constants are constrained. Then we
solve the coupled integral equation for $\pi\rho$ scattering. The
results for the S-wave cross section clearly reveals the $a_1$ meson
with a broad width. We find the pole position in the second Riemann
sheet as $\sqrt{s_R}=(1\,170.7-i 173.0)$ MeV. 
\vspace{0.5 cm}

\section{General formalism~\label{sec:2}}
We start from the definition of the scattering amplitude expressed as 
\begin{align}
	\mathcal{S}_{fi} = \delta_{fi} -i (2\pi)^4\delta^4(P_f-P_i)
  \mathcal{T}_{fi}, 
\end{align}
where $P_f$ and $P_i$ denote the total four-momenta of the final
and initial state. The formal transition amplitude $\mathcal{T}_{fi}$
is obtained from the Bethe-Salpeter (BS) equation with the
coupled-channel formalism employed: 
\begin{align}
	\mathcal{T}_{fi} (p',p;s) &= \mathcal{V}_{fi}(p',p;s) + 
	\frac{1}{(2\pi)^4}\int d^4q \mathcal{V}_{fk}(p',q;s) 
	\mathcal{G}_k(q;s) 
	\mathcal{T}_{ki}(q,p;s), 
\end{align}
where $s$ is the square of the total energy. $p$, $p'$ and $q$ stand
respectively for the four-momenta of the initial, final and
intermediate mesons in the center of mass (CM) frame. The indices 
$i$ and $f$ represent the initial and final meson channels, and $k$
designates the intermediate state in the coupled-channel formalism.  
Since it is rather complicated to deal with the BS equation, we use
its three-dimensional reduction, which is not unique. In the current 
work, we utilize the Blankenbecler-Sugar (BbS)
equation~\cite{Blankenbecler:1965gx, Aaron:1968aoz} that preserves  
the unitarity of two-body interaction for all energies and keeps 
Lorentz invariance. It is convenient to introduce the coupled-channel
formalism and is expressed as 
\begin{align}
	\mathcal{T}_{fi} (\mathbf{p}',\mathbf{p};s) &= \mathcal{V}_{fi}
	(\mathbf{p}',\mathbf{p};s) 
	+\frac{1}{(2\pi)^3}\int \frac{d^3q}{2E_{k1}(\mathbf{q})E_{k2}
	(\mathbf{q})}\mathcal{V}_{fk}(\mathbf{p}',\mathbf{q};s)\frac{E_k
	(\mathbf{q})}{s-E_k^2(\mathbf{q})} 
	\mathcal{T}_{ki}(\mathbf{q},\mathbf{p};s),
	\label{eq:BS-3d}
\end{align}
where $E_{ki} = (\mathbf{q}^2+m_{ki})^{1/2}$ and
$E_k=E_{k1}+E_{k2}$. The zeroth component of the momenta is  
determined by the propagator $\mathcal{G}_k$ given as
$q_0=(E_{k1}-E_{k2})/2$. 

Since we are mainly interested in the $a_1$ meson, we need to consider
only the two channels: $\pi\rho$ and $K\bar{K}^*$ ($\bar{K} K^*$)
ones. Other channels such as the $\pi\omega$ and $\pi\phi$ do not
contribute to the production of the $a_1$ meson. In the coupled-channel
formalism, the kernel $\mathcal{V}_{fi}$ in Eq.~\eqref{eq:BS-3d} is
expressed as  
\begin{align}
	\mathcal{V}_{fi} = \begin{pmatrix}
		\mathcal{V}_{\pi\rho\to\pi\rho} &
                \mathcal{V}_{K\bar{K}^*\to\pi\rho} \\ 
		\mathcal{V}_{\pi\rho\to K\bar{K}^*} &
                \mathcal{V}_{K\bar{K}^*\to K\bar{K}^*}
	\end{pmatrix},
	\label{eq:potfi}
\end{align}
where the off-diagonal part of $\mathcal{V}_{fi}$ contains the
transition from $\pi\rho\to K\bar{K}^*\,(\bar{K}K^*)$. 
Since $K$ and $K^*$ have no definite $G$-parity, we need to combine
the $K\bar{K}^*$ and $\bar{K}K^*$ states, which 
gives a state with the definite $G$ parity:
\begin{align}
	|K\bar{K}^* (\pm)\rangle = \frac{1}{\sqrt{2}}\left(|K\bar{K}^* 
	\rangle \pm|\bar{K}K^* \rangle\right).
\end{align}
Note that we consider only the negative one because $\pi\rho$ has 
negative $G$-parity. We will see later that the $K\bar{K}^*$ channel
with the positive $G$-parity decouples from the $\pi\rho$ channel.

\begin{figure}[htbp]
	\centering
	\includegraphics[scale=0.5]{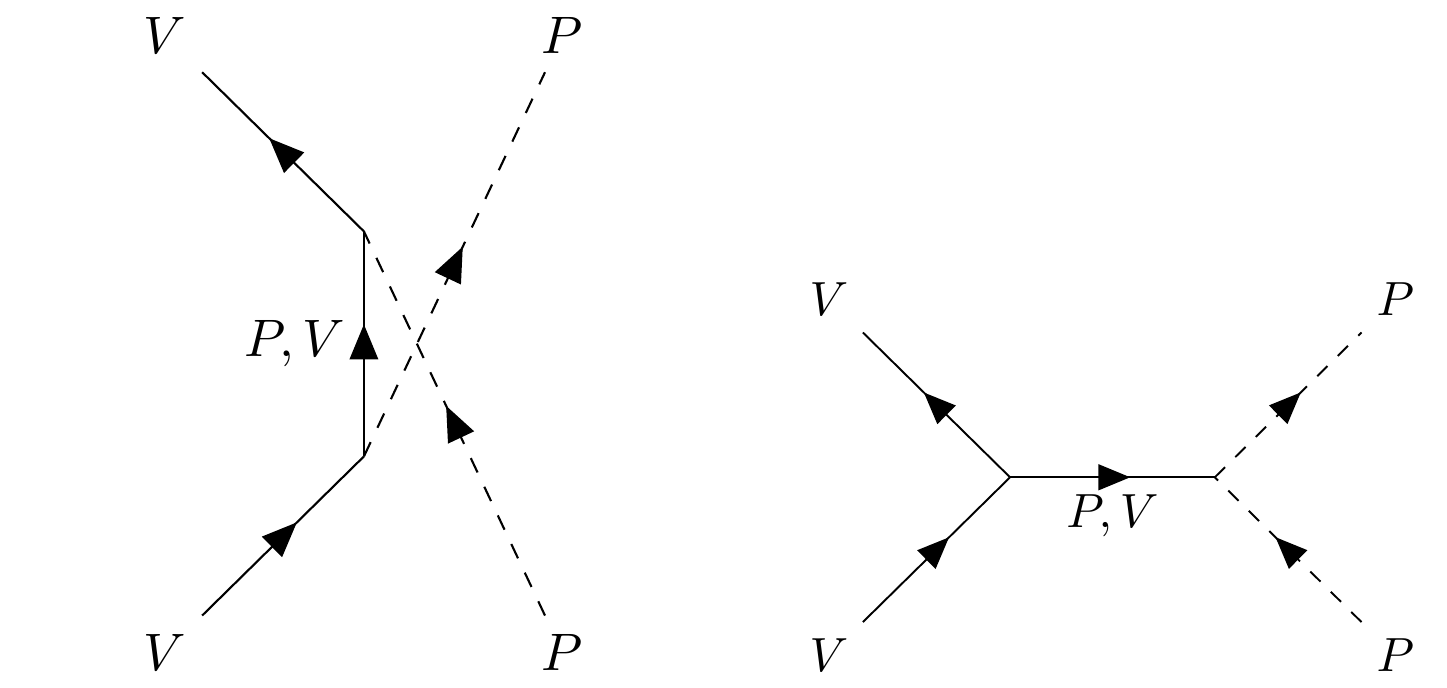}
	\caption{The $u$- (left) and $t$-channels (right) of the
          meson-exchanged diagrams.} 
	\label{fig:1}
\end{figure}

The kernel $\mathcal{V}_{fi}$ in Eq.~\eqref{eq:potfi} is modeled by
meson-exchange diagrams as drawn generically in Fig.~\ref{fig:1}.  
Since the $a_1$ meson will be dynamically generated by the
coupled-channel formalism, we do not include it in the $s$ channel. 
The vertices in the Feynman diagrams are formulated from the SU(3)
symmetric effective Lagrangians given by 
\begin{align}
\mathcal{L}_{PPV} &= \sqrt{2}g_{PPV} \,\mathrm{Tr} 
             \left([P,\partial_\mu P]\,V^\mu\right),\cr 
\mathcal{L}_{VVV} &= -\sqrt{2}g_{VVV} \,
                    \mathrm{Tr}\left((\partial_\mu V_\nu
                    - \partial_\nu V_\mu)V^\mu
                    V^\nu\right),\cr 
\mathcal{L}_{PVV} &=\sqrt{2}\frac{g_{PVV}}{m_V}\,\varepsilon^
{\mu\nu\alpha\beta}
                    \mathrm{Tr}\left(\partial_\mu
                    V_\nu\partial_\alpha V_\beta P\right), 
	\label{eq:su3sym}
\end{align}
where subscripts $V$ and $P$ denote the vector and pseudoscalar mesons
involved in the vertices. $m_V$ represents the mass of the vector
meson. We choose $g_{PPV}=g_{VVV}$ by regarding the vector mesons as
dynamical gauge bosons arising from hidden local gauge
symmetry~\cite{Bando:1984ej, 
  Bando:1985rf}. The values of the coupling constants are taken from
Ref.~\cite{Janssen:1994uf}: $g_{\pi\pi\rho}^2/4\pi=2.84$ and 
$(g_{\pi\rho\omega}^2/4\pi)m_\omega^2=7.5$. These couplings are related to
the $g_{PPV}$ and $g_{PVV}$ by SU(3) symmetric factor as $g_{\pi\pi\rho}=2
g_{PPV}$ and $g_{\pi\rho\omega}m_\omega=2g_{PVV}$. Since 
flavor SU(3) symmetry is broken, the coupling constants vary from the SU(3) 
symmetric case. When it is necessary, we change the values of the coupling 
constants, which are not far from the SU(3) symmetric ones, so that we can 
fit the experimental data. However, we regard the $\phi$-exchange coupling
constant as a free parameter. Its value we have selected differs from
the SU(3) symmetric one by $1~\%$. 

The trace operators in Eq.~\eqref{eq:su3sym} run only over flavor
space. The matrices for the pseudoscalar and vector mesons  
are expressed respectively as
\begin{align}
  P &= \begin{pmatrix}
    \frac{1}{\sqrt{2}} \pi^0+\frac{1}{\sqrt{6}}\eta & \pi^+ & K^+\\
    \pi^- & -\frac{1}{\sqrt{2}} \pi^0+\frac{1}{\sqrt{6}}\eta & K^0\\
    K^- & \bar{K}^0 & -\frac{2}{\sqrt{6}}\eta
  \end{pmatrix},\\
  V_\mu &= \begin{pmatrix}
    \frac{1}{\sqrt{2}} \rho^0_\mu+\frac{1}{\sqrt{2}}\omega_\mu &
    \rho_\mu^+ & K_\mu^{*+}\\
    \rho_\mu^- & -\frac{1}{\sqrt{2}} \rho_\mu^0+\frac{1}{\sqrt{2}}
    \omega_\mu & K_\mu^{*0} \\
    K_\mu^{*-} & \bar{K}^{*0}_\mu & \phi_\mu,
  \end{pmatrix}.
\end{align}
where we employ the standard mixing for the $\omega_1$ and $\omega_8$
components such that the $\omega$ meson contains only the $u$ and $d$
quarks whereas the $\phi$ meson only comprises the strange
quark.
The mixing angle between $\eta$ and $\eta'$ is in the range between  
$-10^\circ$ and $-20^\circ$~\cite{PDG}. It will only lead to $(2-6)~\%$ 
difference in the $g_{K\eta K^*}$ coupling constant, so that $\eta'$
exchange will provide maximum $13~\%$ of the $\eta$-exchange contribution to
the $K\bar{K}^*\to K\bar{K}^*$ potential. Moreover, the contribution
of $\eta$ exchange to the potential is tiny. Therefore, for
simplicity, we ignore $\eta'$ meson exchange in the current study. The
flavor part of Eq.~\eqref{eq:su3sym} can be evaluated in the isospin
bases and yield factors labeled as IS listed in the fourth column of
Table~\ref{tab:1}. 
\begin{table}[htbp]
 \caption{The factor IS and cut-off $\Lambda$ for all possible
 exchange diagram for each reaction. Note that the value
 inside the parentheses is given for the conjugate state of
 $K\bar{K}^*$ and $m$ is the exchange mass.} 
 \label{tab:1}
 \centering\begin{tabular}{|l|c|c|r|c|}
 \hline		
 Reaction & Exchange & Type & IS & $\Lambda-m$ [MeV] \\\hline
 $\pi\rho \to \pi\rho$ & $\pi$ & $u$ & $4$  & $600$\\
           & $\rho$ & $t$ & $-4$  & $600$\\
           & $\omega$ & $u$ & $-4$  & $600$\\
 $\pi\rho \to K\bar{K}^*(\bar{K}K^*)$ & $K$ & $u$ & $-2 (2)$ & $700$\\
           & $K^*$ & $t$ & $2 (-2)$ & $750$\\
 $K\bar{K}^* \to K\bar{K}^*$ & $\rho$ & $t$ & $1$ & $600$\\
           & $\omega$ & $t$ & $-1$ & $600$\\
           & $\phi$ & $t$ & $-2$ & $1400$\\
 $K\bar{K}^* \to \bar{K}K^*$ & $\pi$ & $u$ & $1$ & $600$\\
           & $\eta$ & $u$ & $-3$ & $600$\\
           & $\rho$ & $u$ & $-1$ & $600$\\
           & $\omega$ & $u$ & $1$ & $600$\\
           & $\phi$ & $u$ & $2$ & $1400$\\\hline
	\end{tabular}
\end{table}
Note that the IS factor contains both the SU(3) symmetric factor and
isospin one. From Table~\ref{tab:1}, we find it obvious that 
the $K\bar{K}^*$ channel with positive $G$-parity cannot be coupled
to the $\pi\rho$ channel. 

Since the hadron has a finite size, we introduce the form factor at
each vertex. We use the following parametrization for it 
\begin{align}
	F(t) = \left(\frac{n\Lambda^2-m^2}
	{n\Lambda^2-t}
	\right)^n, \hspace{1 cm}
	F(u) = \left(\frac{n\Lambda^2-m^2}
	{n\Lambda^2-u}
	\right)^n,
\end{align}
where $m$ denotes the mass of the exchange particle. $n$ is determined
by the power of the momentum in the vertex. For example,   
we take $n=1$ for the VPP vertex whereas we choose $n=2$ for the VVP
one. Though the cut-off masses $\Lambda$ are free parameters, we
reduce the uncertainties by fixing their values as follows: we add
($600-700$) MeV to the exchange mass. This idea is based on the fact
that a heavier particle has a smaller size~\cite{Kim:2018nqf,
  Kim:2021xpp, Won:2022cyy}. Thus, the value of $\Lambda$ is also
taken to be larger than that of the corresponding meson mass by around 
($600-700$) MeV. To fit the data, however, we choose a larger value of
the cut-off mass especially for $\phi$ exchange, where its value 
is $1400$ MeV higher than that of the exchanged meson 
mass. In addition, we drop out the energy and angular dependence of
the form factors for the sake of simplicity~\cite{Janssen:1994uf}.    

We obtain the kernels $\mathcal{V}_{fi}$ by summing the 
amplitudes of all possible exchange diagrams listed in Table.~\ref{tab:1}. 
There are only 3 possible diagrams that provide the Feynman amplitudes
as functions of the Mandelstam variables and a type of exchanged
mesons. The amplitudes for the $t$-channel with vector-meson exchange
and for the $u$-channel with pseudoscalar-meson and vector-meson
exchanges are respectively given by
\begin{align}
\mathcal{A}_{V}^{t}(\mathbf{p}',\mathbf{p}) &= \mathrm{IS}\,g^2_{PPV}\,
F^2(t)\,\left(p_2+p_4\right)^\mu\,\left(g_{\mu\nu}-\frac{1}{m_V^2}
(p_1-p_3)_\mu(p_1-p_3)_\nu\right)\mathcal{P}(t) \cr
&\;\;\;\;\times\left[(2p_1-p_3)\cdot\epsilon^*\epsilon^\nu+(2p_3-p_1)
\cdot\epsilon\epsilon^{*\nu}-\epsilon\cdot\epsilon^*(p_1+p_3)^\nu\right],\\ 
\mathcal{A}_{P}^{u}(\mathbf{p}',\mathbf{p}) &= -\mathrm{IS}\,g^2_{PPV}\,
 F^2(u)\,\left(2p_2-p_3\right)\cdot\epsilon^*\,\mathcal{P}(u)\,\left
(2p_4-p_1\right)\cdot\epsilon,\\
\mathcal{A}_{V}^{u}(\mathbf{p}',\mathbf{p}) &= -\mathrm{IS}\,\frac{g^2_
{PVV}}{m_V^2}\,F^2(u)\,\varepsilon_{\mu\nu\alpha\beta}\,p_3^\mu\epsilon^
{*\nu}(p_3-p_2)^\alpha\,g^{\beta\delta}\mathcal{P}(u)\,\varepsilon_
{\gamma\sigma\eta\delta}p_1^\gamma\epsilon^\sigma(p_1-p_4)^\eta,
\end{align}
where $p_1(\mathbf{p})$ and $p_2(\mathbf{p})$ respectively denote the 
four momenta of the initial vector and pseudoscalar mesons whereas 
$p_3(\mathbf{p}')$  and $p_4(\mathbf{p}')$ are those of the final
vector and pseudoscalar mesons respectively. The polarization vectors 
of the initial and final vector mesons are respectively labeled as
$\epsilon(\mathbf{p})$ and $\epsilon^*(\mathbf{p}')$. As for the
propagators of the exchange mesons, we utilize  
the static ones, following Ref.~\cite{Janssen:1994uf}. 

Since the $a_1$ meson arises from the $S$-wave transition amplitude, we
carry out the partial wave decomposition of the 
kernel and transition amplitudes. The partial-wave helicity amplitudes
can be obtained by projecting the amplitudes onto the total angular
momentum $J$
\begin{align}
	\mathcal{T}^{J(fi)}_{\lambda'\lambda} (\mathrm{p}',\mathrm{p}) &= 
	\mathcal{V}^{J(fi)}_{
	\lambda'\lambda} (\mathrm{p}',\mathrm{p}) + \frac{1}{(2\pi)^3}
	\sum_{k,\lambda_k}\int
	\frac{\mathrm{q}^2d\mathrm{q}}{2E_{k1}(\mathrm{q})E_{k2}(
	\mathrm{q})} \mathcal{V}^{J(fk)}_{\lambda'\lambda_k}(\mathrm{p}',
	\mathrm{q})\frac{E_k(\mathrm{q})}{
	s-E_k^2(\mathrm{q})} \mathcal{T}^{J(ki)}_{\lambda_k\lambda}
	(\mathrm{q},\mathrm{p}),
	\label{eq:BS-1d}
\end{align}
where $\lambda'$, $\lambda$, and $\lambda_k$ denote the helicities of 
the final ($f$), initial ($i$) and intermediate ($k$) states, 
respectively. The partial-wave kernel amplitudes can be
expressed as 
\begin{equation}
	\mathcal{V}^{J(fi)}_{\lambda'\lambda}(\mathrm{p}',\mathrm{p}) = 
	2\pi \int d(
	\cos\theta) \,d^{J}_{\lambda'\lambda}(\theta)\,\mathcal{V}^{fi}_{
	\lambda'\lambda}(\mathrm{p}',\mathrm{p},\theta),
	\label{eq:pwd}
\end{equation} 
where $\theta$ denotes the scattering angle and $d^{J}_{\lambda'\lambda}
(\theta)$ stand for the matrix elements of the Wigner $D$
functions. The partial-wave $\mathcal{T}$ amplitude is also expressed
in a similar manner. 

The partial-wave coupled integral equation in Eq.~\eqref{eq:BS-1d} is  
solved numerically after we regularize the singularity arising from the
two-body meson propagator $\mathcal{G}$. We build the matrix
$\mathcal{V}$ in momentum space, including both the $\pi\rho$ and 
$K\bar{K}^*$ channels. Then $\mathcal{T}$ matrix can be derived by the 
Haftel-Tabakin's method of the matrix inversion~\cite{Haftel:1970zz} 
\begin{align}
	\mathcal{T} =
  \left(1-\mathcal{V}\tilde{\mathcal{G}}\right)^{-1}  
	\mathcal{V}.
\end{align}
It is convenient to write the $\mathcal{T}$ matrix in the particle  
basis~\cite{Machleidt:1987hj}. Thus, we define $\mathcal{T}_{IJL}$ as
the $\mathcal{T}$ matrix for a given total isospin $I$, total angular
momentum $J$, and orbital angular momentum $L$. 

\section{$a_1$ meson as a $K\bar{K}^*$ molecular state~\label{sec:3}}
\begin{figure}[htbp]
 \centering
 \includegraphics[scale=0.55]{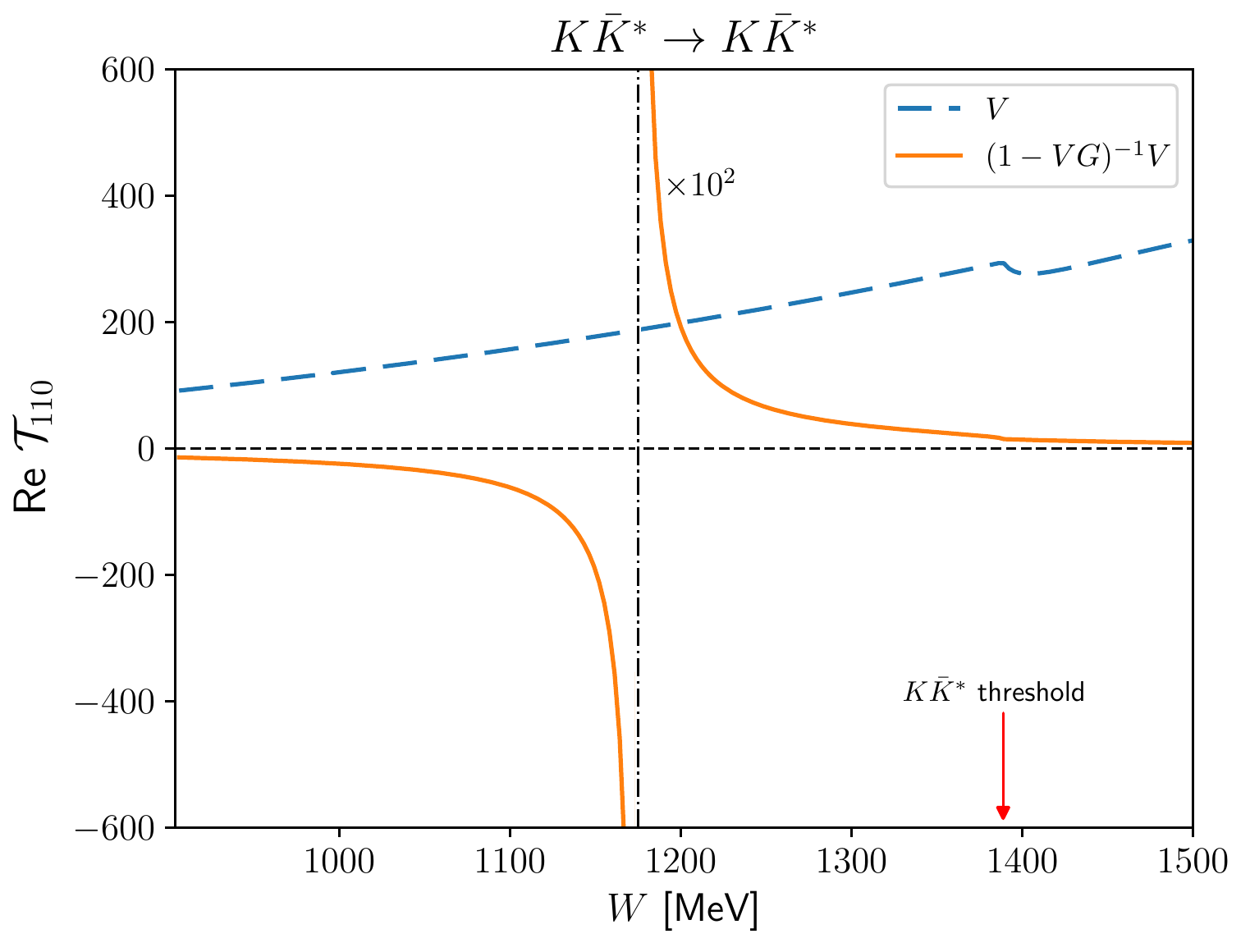}
 \caption{Real part of $\mathcal{T}_{110}$ in the
  $K\bar{K}^*\to K\bar{K}^*$ reaction as a function of
  energy. Here we label the kernel amplitude
    $\mathcal{V}_{K\bar{K}^*\to K\bar{K}^*}$ as $\mathcal{V}$. The
  dashed line depicts $\mathcal{V}$, whereas
  the solid curve draws $(1-\mathcal{VG})^{-1}\mathcal{V}$.} 
 \label{fig:2}
\end{figure}

We discuss now how the $a_1$ meson can be dynamically generated by the
coupled channel formalism. In Fig.~\ref{fig:2}, we draw the real part 
of $\mathcal{T}_{110}$ in the $K\bar{K}^*\to K\bar{K}^*$ channel,
which corresponds to the quantum numbers of the $a_1$ resonance. 
Here we will only consider the $K\bar{K}^*$ single channel to examine
how the resonance behavior arises from the integral equation.
The kernel amplitude $\mathcal{V}_{K\bar{K}^*\to K\bar{K}^*}$ itself
does not show any resonance behavior, which is depicted as the dashed
line. On the other hand, the full transition amplitude generates the
singularity below the $K\bar{K}^*$ threshold energy after the integral
equation is solved.  The pole is positioned on the real energy axis. 
This singularity occurs from the strong attraction of the
$t$-channel exchange. Specifically, this attractive potential comes
from the $\phi$ exchange diagram since the $\rho$ and $\omega$
contribution cancelled each other as we can see from their IS
factor. This singular behavior in $\mathcal{T}$ is responsible for
creating the $a_1$ meson in the $\pi\rho\to\pi\rho$ reaction. The
remarkable point is that it only appears below the $K\bar{K}^*$
threshold. Once we introduce the $\pi\rho$ channel and make it coupled
to the $K\bar{K}^*$ one, the pole moves to the second Riemann sheet in
the complex energy plane. This indicates that the finite width of the $a_1$
resonance is caused by the coupling of the $K\bar{K}^*$ channel with
the $\pi\rho$ one.  

\begin{figure}[htbp]
 \centering
 \includegraphics[scale=0.55]{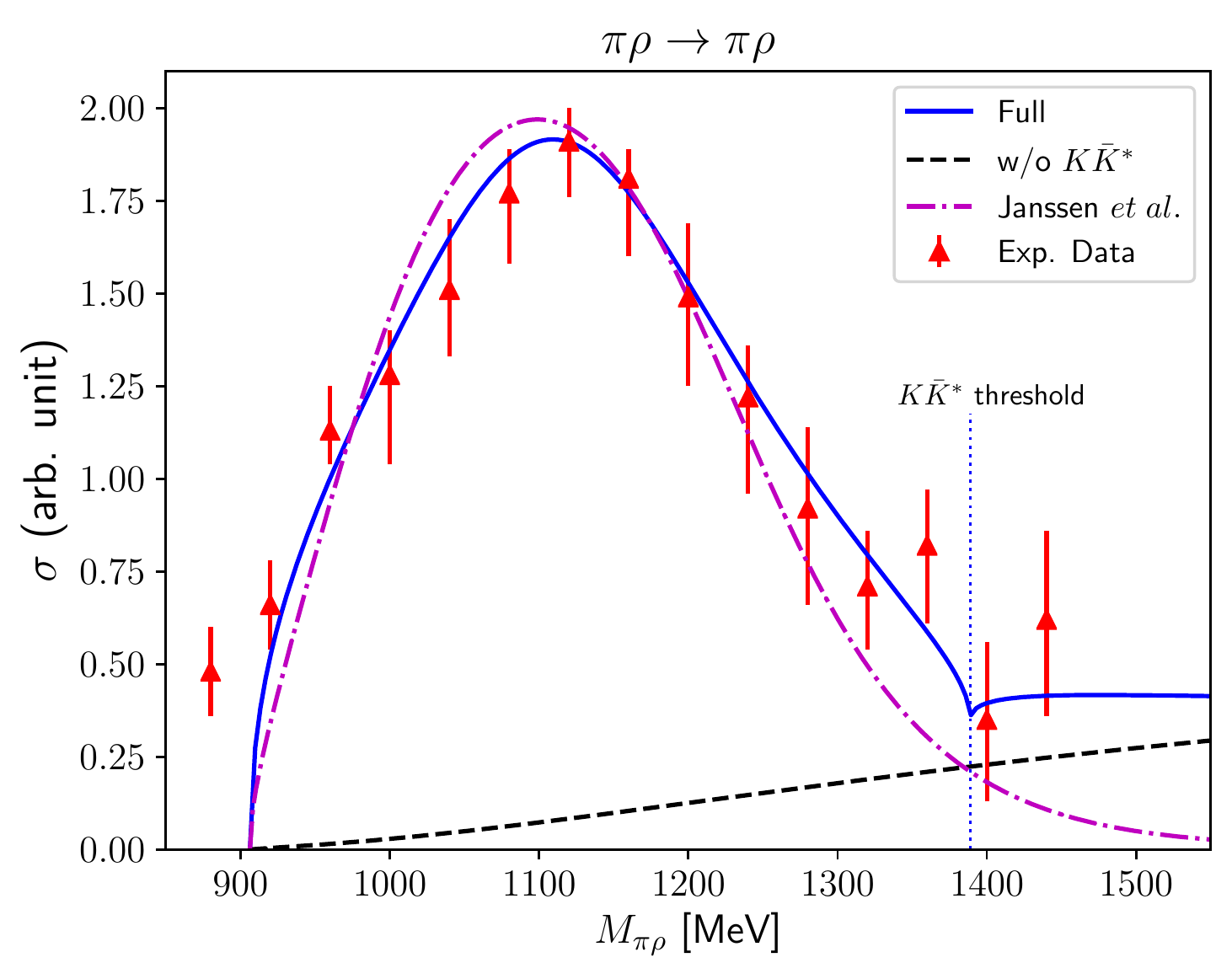}
 \caption{Comparison of the  $\pi\rho\to\pi\rho$ total cross
   section for $IJL=110$ as a function of $\pi\rho$ invariant
   mass with that from Ref.~\cite{Janssen:1994uf}. 
   The solid curve draws the result from the present work whereas the 
   dot-dashed one depicts that from Ref.~\cite{Janssen:1994uf}. 
   The dashed line exhibits the result with the $K\bar{K}^*$ channel
   turned off. The experimental data are taken from
   Ref.~\cite{Dankowych:1981ks}.}  
 \label{fig:3}
\end{figure}
To compared the experimental data~\cite{Dankowych:1981ks} given in an
arbitrary unit, one can take the total cross section to be 
\begin{align}
	\sigma \equiv \sigma_{\pi\rho}\left(t=m_\rho^2, M_{\pi\rho}\right)
	 = -C\, \mathrm{Im}[\mathcal{T}_{\pi\rho}(M_{\pi\rho})],
\end{align}
where $C$ is the constant to match the data to the results from a 
theoretical model. The full transition amplitude
$\mathcal{T}_{\pi\rho}$ is obtained by solving the integral equation 
in the coupled-channel formalism, given in Eq.~\eqref{eq:BS-3d}. 
Figure~\ref{fig:3} shows the result for $\sigma$ 
as a function of the $\pi\rho$ invariant mass.
The solid curve draws the current result whereas the dot-dashed one
corresponds to that from Ref.~\cite{Janssen:1994uf} in which the $a_1$
meson was explicitly introduced as a $s$-channel pole diagram with the
$\pi\rho$ channel only considered. The result from
Ref.~\cite{Janssen:1994uf} shows a symmetric shape, since the $a_1$
pole diagram governs it. On the other hand, the present result reveals
a dynamical feature of the $a_1$ resonance. The physical $a_1$ meson
is generated only by coupling the $\pi\rho$ channel with the
$K\bar{K}^*$ one. It describes well the experimental data on 
$\pi\rho$ scattering near the $\pi\rho$ threshold. The S-wave total
cross section rapidly increases and reaches the maximum value at
around $1\,100$ MeV.  While the result from Ref.~\cite{Janssen:1994uf}
falls off rapidly, so that it is underestimated in the vicinity of the
$K\bar{K}^*$ threshold, the present result decreases less rapidly. It 
explains well the data even near the $K\bar{K}^*$ threshold. If we
turn off the $K\bar {K}^*$ channel, the result cannot yield the
$a_1$(1260) resonance structure at all. Thus, the $K\bar{K}^*$ channel
plays a critical role in generating the $a_1$ resonance. A similar
situation can be found in the case of the $f_0(980)$ meson. In 
Ref.~\cite{Lohse:1990ew}, $\pi\pi$ scattering was investigated within
the meson-exchange model, where the $f_0(980)$ resonance can only
appear when the $K\bar{K}$ channel was included. A similar feature was
also observed in Ref.~\cite{Oller:1997ti}. So, the $f_0(980)$ meson is
often considered as a $K\bar{K}$ molecular state. Similarly, the
$a_1(1260)$ resonance can be called a $K\bar{K}^*$ molecular state.   

To scrutinize the $a_1(1260)$ resonance based on the current work, we 
evaluate the pole position for this resonance in the second Riemann
sheet and coupling strength at the pole position. We utilize 
the analytic continuation method~\cite{Suzuki:2008rp} to
determine the scattering matrix on the complex energy plane of total
energy. We locate the pole position $a_1$ at 
$\sqrt{s_R} = (1170.7 - i 173.0)$ MeV in the complex
energy plane. Since there is no other resonance nearby, we can
determine clearly its position. From the pole position we found that 
the width of the $a_1$ meson is in agreement with that
from Ref.~\cite{Dankowych:1981ks}, where $\Gamma = (380\pm
100)$ MeV was obtained by using the Bowler model fit. Experimentally,
the width of the $a_1(1260)$ meson is given in the wide range:
$\Gamma_{a_1}=(250-600)$ MeV~\cite{PDG}. The present work provides
almost the center value of $\Gamma_{a_1}\approx 350$ MeV, compared to
the PDG data.  

\begin{figure}[htbp]
	\centering
	\includegraphics[scale=0.4]{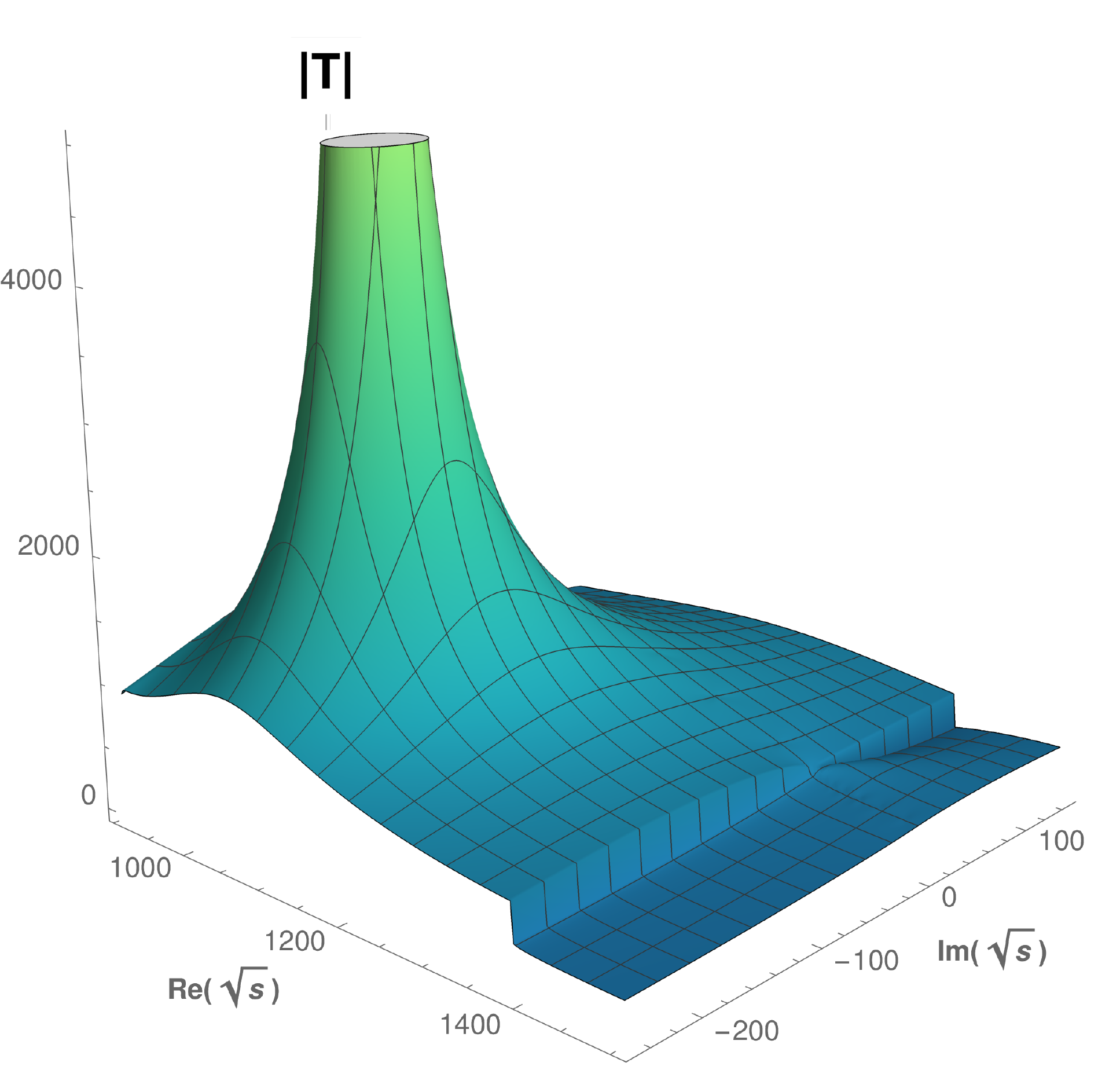}
	\caption{The 3D plot of absolute $T$ matrix as a function of 
	complex energy.}
	\label{fig:4}
\end{figure}
To see the $a_1$ resonance structure more explicitly, 
we also present the 3D plot of the $|T|$ in the complex energy plane,
as illustrated in Fig.~\ref{fig:4}. To derive the coupling strengths
of the $a_1$ meson coupled to the $\pi\rho$ and $K\bar{K}^*$ channels,
we derive it from the residue of the transition amplitude defined as 
$\mathcal{R}_{a,b}$: 
\begin{align}
 \mathcal{R}_{a,b} =\lim_{s\to s_R}\left(s-s_R\right) \mathcal{T}_{a,b}/4\pi .
 \end{align}
We divide the partial-wave component of the $T$-matrix with the factor 
$4\pi$ since we use different definition of partial wave expansion in 
Eq.~\eqref{eq:pwd}. The coupling strengths are defined as the square
root of the residue of the transition amplitude, so that we obtain them as 
\begin{align}
	g_{\pi\rho}^{a_1} &= \sqrt{\mathcal{R}_{\pi\rho,\pi\rho}^{a_1}} = 
	(5.75 - i1.35)\, [\mathrm{GeV}],\\
	g_{K\bar{K}^*}^{a_1} &= \sqrt{\mathcal{R}_{K\bar{K}^*,K\bar{K}^*}^
	{a_1}} =(12.37 - i 2.31)\, [\mathrm{GeV}].
\end{align}
Note that we choose the positive signs for both coupling strengths,
since we are not able to determine them. 

Finally but not least, we want briefly to mention about the
compositeness of the $a_1$ meson. Weinberg shows the possibility to
quantify the nature of the bound state through the renormalization 
constant~\cite{Weinberg:1962hj}. It is possible to extend this idea to
the virtual and resonance state. In Ref.~\cite{Hyodo:2013nka}, for
instance, the compositeness of an unstable resonance was studied by
examining the coupling strength and the derivative of a two-body loop
propagator evaluated in the second Riemann sheet where the resonance's
pole is located. In Ref.~\cite{Guo:2015daa} the positive definite
compositeness of the resonance was studied. However, it is valid to
examine the compositeness only when the threshold energy is less than
the mass of a resonance. Therefore, it is rather complicated to
calculate the compositeness to prove that $K$ and $K^*$
dominate in the $a_1$ since the $K\bar{K}^*$ threshold energy is
larger than the $a_1$ mass in the present work.

\vspace{1em}
\section{Summary and conclusions~\label{sec:5}} 
In the present work, we studied $\pi\rho$ scattering based on the
meson-exchange model, focusing on the $a_1(1260)$ resonance appearing
in the $S$-wave total cross section. We showed that the coupled-channel
formalism including the $\pi\rho$ and $K\bar{K}^*$ channels generated
the $a_1$ meson dynamically. The $K\bar{K}^*$ channel plays an
essential role in producing the $a_1$ meson. We solved the integral
equation for $K\bar{K}^*$ scattering and found a pole on the real
energy axis. Once we introduced the $\pi\rho$ channel and coupled it
to the $K\bar{K}^*$ channel, we observed that the $a_1$ resonance
arises dynamically in $\pi\rho$ scattering. We obtained the pole
position of the $a_1$ resonance at $\sqrt{s_R} = (1170.7 - i 173.0)$
MeV. The present result is much better than the one in the previous
study, where the $a_1$ meson was introduced explicitly in the
$s$-channel pole diagram with the $\pi\rho$ single channel considered
only. We also derived the coupling strengths $g_{\pi\rho}^{a_1}$ and
$g_{K\bar{K}^*}^{a_1}$ from the residue of the transition
amplitude. These results imply that the dynamically generated $a_1$
meson may be interpreted as a $K\bar{K}^*$ molecular state.

\vspace{0.5cm}
\begin{acknowledgments}
The present work was supported by Basic Science Research Program
through the National Research Foundation of Korea funded by the 
Korean government (Ministry of Education, Science and Technology,
MEST), Grant-No. 2021R1A2C2093368 and 2018R1A5A1025563. 
\end{acknowledgments}

\bibliography{a1Resonance}

\end{document}